\documentstyle[aps,prl,multicol,epsf]{revtex}

\begin{document}
\draft
\preprint{HKBU-CNS-9801}
\title{
Nonadiabatic Geometric Phase and Hannay Angle: A Squeezed State
Approach
}
\author{
Jie Liu $^{1,2}$,
Bambi Hu $^{1,3}$, and Baowen Li$^1$ 
}
\address{$^1$ Department of Physics and Centre for Nonlinear Studies,
Hong Kong Baptist University, Hong Kong, China\\
$^2$ Institute of Applied Physics and
Computational Mathematics, P.O.Box.8009,  100088 Beijing, China\\
$^3$ Department of Physics, University of Houston, Houston, TX 77204 
}
\date{\today}
\maketitle
\begin{abstract}
The geometric phases of the cyclic states of a generalized 
harmonic oscillator with nonadiabatic time-periodic parameters are 
discussed in the framework of squeezed  state. 
A class of cyclic states
are  expressed as a superposition of an infinte
number of squeezed  states.
Then, their geometric phases are  obtained explicitly and found to be
 $-(n+1/2)$ times the classical nonadiabatic
Hannay angle.
It is shown that the analysis based on squeezed state approach
provide a clear
picture of the geometric meaning of the quantal phase.
\end{abstract}

\begin{multicols}{2}
Berry phase\cite{Berry84}, which reveals the gauge structure associated
with a phase shift in adiabatic processes in quantum mechanics, has
attracted great theoretical interests and has been repeatedly corroborated
by experiments (see e.g.\cite{SW89}). This quantum adiabatic phase has a
classical analog -Hannay angle \cite{HB85}.  The relaxation of the
adiabatic approximation is an important step \cite{AA87,BH88}. Aharanov
and Anandan \cite{AA87} studied the phase associated with a cyclic
evolution in quantum mechanics (which occurs when a state returns to its
initial condition), and shown that the phase is a geometric property of
the curve in the projective Hilbert space which is naturally associated
with the motion. 

The significance of Aharanov and Anandan's generalization are twofold. On
the one hand, the cyclic evolution of a physical system is of most
interest in physics both experimentally and theoretically. On the other
hand, the universal existence of the cyclic evolution is guaranteed for
any quantum system.  This can be easily recognized by considering the
eigenvectors of the unitary evolution operator for a quantum system.  An
explicit example is a time-periodic Hamiltonian system where the Floquet
theorem applies. The eigenfunctions of the Floquet operator, which are
so-called the Bloch wave functions in the condensed matter physics, are
obviously cyclic solutions and of great interest in physics. Unlike the
adiabatic case, however, in the nonadiabatic case, calculating the
eigenvectors and extracting the nonadiabatic geometric phase from the
quasi-energy term for a time-dependent Hamiltonan is far from trival,
except for such a special example as the spin particle in a magnetic
field.  Recent works of Ge and Child\cite{GC97} made a step further in
this direction. They found a {\em special} cyclic state of Gaussian wave
packet's form for a generalized harmonic oscillator. The nonadiabatic
geometric phase is explicitly calculated and found to be one half of the
classical nonadiabatic Hannay angle. 

In this letter, we would like to study the nonadiabatic geometric
phase of the {\em general} cyclic evolutions of the generalized harmonic
oscillator. To this end, an alternative way - squeezed state approach will
be used. In particular, we shall construct a class of quantum states based
on a superposition of an infinite number of squeezed states.
We find that the condition for them to be cyclic evolutions is
nothing but a quantization rule without Maslov-Morse correction.
 The
nonadiabatic geometric phases are obtained {\it explicitly}, and found
to be related to the classical Hannay angle by a factor $n+1/2$. 
Furthermore, the quantum phase can be interpreted as a sum of the area
difference on the expectation value plane through a cannonical
transformation and the area on the quantum fluctuation plane swept out by
a periodic orbit. This interpretation gives a unified picture of the
geometric meaning of the quantal phase for the adiabatic and nonadiabatic
case. 
 
Squeezed state approach has been successfully applied in many branches of
physics such as quantum optics, high energy physics and condensed matter
physics. Recent years have witnessed a growing application of squeezed
state to study the chaotic dynamical systems\cite{PS94,ZF95,LS97,HLLZ98}.
In this letter, we shall employ this approach to discuss geometric phase
and Hannay angle for a generalized harmonic oscillator. 
An apparent reason for this choice is that this system admits the squeezed
state as an exact solution. The squeezed state
approach\cite{JK79,ZFG90,TF91} starts from the time-dependent variational
principle (TDVP) formulation,

\begin{equation}   
\delta \int dt\langle\Phi,t|i\hbar\frac{\partial}{\partial t} -
\hat H |\Phi,t\rangle = 0.   
\end{equation}

Variation w.r.t $\langle\Phi,t|$ and $|\Phi,t\rangle$ gives rise to the
Schr\"odinger equation and its complex conjugate, respectively. 
The squeezed state is chosen as  the trial wave function, which   is defined
by the ordinary harmonic oscillator displacement operator acting on
a squeezed vacuum state $|0\rangle$:

$$|\Psi\rangle = \exp\left(\alpha \hat a^{+} - \alpha^{*} \hat 
a\right)|\phi\rangle,$$
\begin{equation}
|\phi\rangle = \exp\left(\frac 1 2 (\beta{\hat a^{+2}} - \beta {\hat 
a}^2)\right)|0\rangle.
\end{equation}
${\hat a}^{+}$ and ${\hat a}$ are boson creation and annihilation operator
which satisfy the canonical commutation relation: $[\hat a,{\hat a}^+] =1$. 

From the TDVP, we can obtain the      
dynamical equations for the expectation values $(q,p)$ and the    
quantum fluctuations
$ 
\Delta p^2 \equiv \langle\Psi,t|(\hat p-p)^2|\Psi,t
\rangle=\hbar(\frac{1}{4G}+4\Pi^2G),
\Delta q^2 \equiv \langle\Psi,t|(\hat q-q)^2|\Psi,t\rangle=\hbar G,
$
\begin{eqnarray}
\dot q =\frac{\partial H_{eff}}{\partial p},\qquad
\dot p = -\frac{\partial H_{eff}}{\partial q},\nonumber\\
\hbar \dot G = \frac{\partial H_{eff}}{\partial \Pi},\qquad
\hbar \dot \Pi = -\frac{\partial H_{eff}}{\partial G},
\end{eqnarray}
where the dot denotes the time derivative.  The  effective 
Hamiltonian  $H_{eff}$
is  defined on the extended space $(q,p,G,\Pi)$, taking the form      
$H_{eff} = \langle\Psi,t|\hat H|\Psi,t\rangle$.

The time-dependent variational principle leaves an ambiguity of a
time-dependent phase $\lambda (t)$, which can be fixed with the aid of the
Schr\"odinger equation,
      
\begin{equation}      
\dot \lambda(t) = \langle\Psi,t|i\frac{\partial}{\partial t}|
\Psi,t\rangle      
-\frac{1}{\hbar}\langle\Psi,t|\hat H|\Psi,t\rangle.      
\end{equation}

This phase is well defined for general {\it nonadiabatic} and 
{\it noncyclic} evolution of a squeezed state.
It represents a phase change of the squeezed state during
a time-evolution.
Obviously, the phase consists of two parts. The meaning of the second part is 
obvious: a measure of the time of evolution. It is the {\it dynamical} 
phase  and can be rewritten as,          

\begin{equation}
\lambda_D(t) = -\frac{1}{\hbar}\int_0^t H_{eff} dt.
\end{equation}
The first part can be viewed as a difference of the {\it total} phase 
and the {\it dynamical} phase. We call it  {\it geometric} phase since it
just is the Aharanov-Anandan's phase
for  the case of cyclic evolution. From the expression
of the squeezed state, the geometric phase  is equal to 

\begin{equation}
\lambda_G(t) =\int_0^t \left(\frac{1}{2\hbar}(p\dot q - q\dot p) -\dot\Pi 
G\right) dt. \end{equation}
 
It is clear that  the evolution of
expectation values $(q,p)$ 
as well as the evolution of the quantum fluctuations $(G,\Pi)$ contribute
 to the geometric phase. The contribution from the former one is
explicitly $\hbar$ dependent, while the contribution from quantum fluctuation
is $\hbar$ independent. 
For the case of cyclic evolution of squeezed state the
 quantal phase is equal to a sum of the projective areas on the 
coordinates plane $(q,p)$  and fluctuation plane $(G,\Pi)$
 swept out by a periodic orbit
of the effective Hamiltonian.

The Hamiltonian of the generalized harmonic oscillator 
takes the form,
        
\begin{equation}       
\hat H(q,p,t) =        
\frac 1 2 \left(a(t)\hat q^2 + b(t)\hat p^2       
+ c(t)(\hat q \hat p + \hat p \hat q)\right),       
\label{GHO}
\end{equation}      
where real parameter ${a(t),b(t),c(t)}$ are time-periodic functions        
with common period $T$.
Our discussions are restricted to the elliptic case, namely,  $a(t)b(t) > 
c^2(t)$.

Applying the squeezed state to this system, from Eq. (\ref{GHO})
one obtains an effective Hamiltonian in the extended phase space        
$(q,p;G,\Pi)$,
        
\begin{equation}        
H_{eff}(q,p;G,\Pi;t) = H_{cl}(q,p,t) + \hbar H_{fl}(G,\Pi,t),        
\end{equation}
where  
\begin{equation}
H_{cl} =         
\frac 1 2 \left(a(t) q^2 + b(t) p^2 + 2 c(t) q p\right),
\end{equation}
describes the motion of the expectation values;
\begin{equation}
H_{fl} = \frac 1 2 \left(a(t)G + b(t)(\frac{1}{4G} +4\Pi^2G) +       
4c(t)G\Pi\right),
\end{equation}
depicts the evolution of the quantum fluctuations.

Starting from this effective Hamiltonian, it is easy to analyse the
dynamical properties. The motions of both degree of freedom are decoupled.
In the fluctuation plane $(G,\Pi)$, whole motions are restricted on the
invariant tori except for a unique T-periodic solution
denoted by $(G_p(t),\Pi_p(t))$. The Hamiltonian $H_{cl}$  which 
describes the motion of the expectation values $(q,p)$
is identical to the 
classical version of the system  (\ref{GHO}). 
The $(q=0,p=0)$ is obviously  a fixed point. 
Other motions  are  quasi-periodic
trajectories confined on the tori. 
Through a canonical transformation, $q=q(\bar I,\bar\phi,t),
p=p(\bar I,\bar\phi,t)$,
the Hamiltonian $H_{cl}(q,p,t)$
can be transformed to a new Hamiltonian $\bar H(\bar I,t)$ which 
does not contain  angle variable $\bar\phi$.
Its solution is described by $\bar I =\bar I_0;
\bar\phi(t) = \bar\phi_0 + \int_0^t
\frac{\partial \bar H(\bar I_0, t)}
{\partial \bar I_0 } dt .$
For this canonical transformation is explicitly time dependent,
the new Hamiltonian $\bar H$ differs from the
old one $H_{cl}$ both in value and in functional form. 
Thus, we introduce   a function $A$ to measure  the difference,

\begin{equation}
A(\bar \phi, \bar I,t) = \bar H(\bar I,t) - H_{cl}\left((\phi(\bar\phi,\bar 
I,t), I (\bar \phi,\bar I,t),t\right).
\end{equation}

Therefore the classical non-adiabatic Hannay angle is

\begin{equation}
\Theta_H =\langle \int_0^T \frac{\partial A}{\partial \bar I}dt \rangle_
{\bar \phi_0},
\label{Hannay}
\end{equation}
where the bracket denotes  averaging around the
invariant torus,
$\langle \cdots\rangle=\frac{1}{2\pi}\int_0^{2\pi}\cdots d\bar\phi_0$.

Now we turn to the quantum system (\ref{GHO}). Since it is a time-periodic
Hamiltonian system, the Floquet theory applies. A unitary time evolution
operator refering to one period T, the so-called Floquet operator $\hat
U(T)$ is worthy of consideration. We can construct a state as a
superposition of infinite number of squeezed states

\begin{equation}
|S_1\rangle = c\int_0^{2\pi}
e^{\frac{i}{\hbar} \bar I_0 \bar\phi_0}|
\bar I_0,\bar\phi_0;G_0,\Pi_0\rangle d\bar\phi_0 ,
\end{equation}
where 
$|\bar I_0,\bar\phi_0;G_0,\Pi_0\rangle$
represents a squeezed state centered at $
q(\bar I_0,\bar \phi_0,t=0)
,p(\bar I_0,\bar \phi_0,t=0)$
with fluctuations $G_0,\Pi_0$; 
The $G_0$ and $\Pi_0$ are chosen on  the unique periodic orbit 
$(G_0=G_p(t=0), \Pi_0=\Pi_p(t=0))$ ; $c$ is a 
normalization constant.

Consider the situation that $\hat U(mT)$ (or $\hat U^m(T)$) acts on the
state $|S_1\rangle$,

\begin{equation}
\hat U(mT)|S_1\rangle = 
c\int_0^{2\pi}e^{\frac{i}{\hbar}\bar I_0 \bar \phi_0}
e^{i\lambda}
|\bar I_0,\bar\phi_0+\bar \phi^m
;G_0,\Pi_0\rangle
d\bar \phi_0,
\end{equation}
where 
$\bar \phi^m=\int_0^{mT}\frac{\partial \bar H(\bar I_0,t)}{\partial \bar 
I_0}dt$, and 
$ 
\lambda = \lambda_D(mT) + \lambda_G(mT).
$
The dynamical part is $
 \lambda_D(mT)=-\frac{1}{\hbar}\int_0^{mT} H_{eff}dt,\quad 
$
and the geometric part
$ \lambda_G(mT) = \frac{1}{\hbar}\int_0^{mT}\frac {1}{2}
(p\dot q -q\dot p)dt
-\int_0^{mT}\dot\Pi_p G_p dt.
$
They can be expressed as,

\begin{equation}
\lambda_D(mT)=
\langle\lambda_D(mT)\rangle_{\bar \phi_0} + \{\lambda_D(mT) \}(\bar\phi_0),
\label{pch1}
\end{equation}
\begin{equation}
\lambda_G(mT)=\langle\lambda_G(mT)\rangle_{\bar \phi_0}
+ \{\lambda_G(mT) \}(\bar\phi_0).
\label{pch2}
\end{equation}
respectively. Where the symbols $\langle \cdots\rangle_{\bar\phi_0}$
denotes the average over the $\bar \phi_0$ as in Eq.(\ref{Hannay});
$\{\cdots\}(\bar\phi_0)$ represent
the terms relating to   $\bar\phi_0$.
Then,

\begin{equation}
\langle\lambda_G(mT)\rangle_{\bar\phi_0}
= \frac{m}{\hbar}\langle\int_0^T \left(\frac 1 2 (p\dot q
-q\dot p)\right)dt\rangle_{\bar\phi_0}
-m\oint G_p d\Pi_p .
\end{equation}

Making variables transformation $\bar\phi_0' = \bar\phi_0 +  \bar \phi^m$, 
we have

$$\hat U(mT)|S_1\rangle = 
c e^{i\lambda_m^1}
\int_{\bar \phi^m}^{2\pi+\bar \phi^m}
e^{\frac{i}{\hbar}\bar I_0\bar\phi_0'}$$
\begin{equation}
e^{i\{\lambda_D(mT)\}(\bar\phi_0')+i\{\lambda_G(mT)\}(\bar\phi_0')}|
\bar I_0,\bar \phi_0';G_0,\Pi_0\rangle d\bar\phi_0' ,
\label{ums1} 
\end{equation}
where
$ \lambda_m^1 = m (\lambda_G^R+\lambda_D^R).
$
The geometrical  part and the dynamical part take the forms as follows,

$$\lambda_G^R 
= \frac{1}{\hbar}
\left(\langle\int_0^T \frac 1 2 (p\dot q
-q\dot p)dt\rangle_{\phi_0}
-\bar I_0\int_0^T\frac{\partial \bar H}{\partial \bar 
I_0} dt\right)$$
\begin{equation}
-\oint G_p d\Pi_p.
\label{Fgphase}
\end{equation}
 \begin{equation}
 \lambda_D^R=
-\frac{1}{\hbar}
\langle\int_0^T H_{eff} dt\rangle_{\phi_0}.
\end{equation}

The integral in Eq.(\ref{ums1}) can be written as
$
\int_{0}^{2\pi}\cdots
+
\int_{2\pi
}^{2\pi+\bar \phi^m}\cdots
-
\int_{0
}^{\bar \phi^m} \cdots.
$
The last two terms will   cancell each other if and only if
$
e^{\frac{i}{\hbar} \bar I_0 2\pi}=1$, which gives rise to

\begin{equation}
\bar I_0 = n\hbar .
\label{quru}
\end{equation}
This is nothing but  the quantization rule without Maslov-Morse correction.

The motion of the expectation values $(q,p)$ confined on the
invariant torus $\bar I_0$ is quasi-periodic. The ergodicity of the motion
guarantees that temporal average is equivalent to the spatial
average supposing that the time is long enough. Then, from the ergodicity
principle, we can choose an
 integer $r$, which is large enough so that
the phase change (see(\ref{pch1}) and (\ref{pch2}))
during the time interval  $rT$ does not relate to
$\bar\phi_0$.
Then,  we construct a state $|S_r\rangle$ like\cite{LHL98},

$$|S_r\rangle = 
|S_1\rangle + \cdots
+ e^{-i\lambda_m^1}\hat U(mT)|S_1\rangle
+ \cdots$$
\begin{equation}
+e^{-i\lambda_{r-1}^1}\hat U((r-1)T)|S_1\rangle.
\label{stater}
\end{equation}
and under the condition (\ref{quru}) 
 we can prove that\cite{LHL98},
\begin{equation}
\hat U(T)|S_r\rangle = e^{i(\lambda_D^R +\lambda_G^R)}|S_r\rangle.
\end{equation}

In fact, the above relation indicates that 
the state $|S_r\rangle$ is an eigenstate of the Floquet
operator, $n$ is the state number. $\lambda_D^R$ and $\lambda_G^R$
is the dynamical and geometric phase relating to the cyclic states
, respectively.

To  see the meaning of the geometric
phase $\lambda_G^R$ expressed by Eq.(\ref{Fgphase}), let us 
consider following differential
2-form which is preserved under  the canonical transformation, i.e.
$
dp\wedge dq -dH\wedge dt = d\bar I\wedge d\bar \phi
-d\bar H\wedge dt.
$
 We rewrite this into another form,

\begin{equation}
dp\wedge dq -   d\bar I\wedge d\bar \phi = -d(\bar H -H_{cl})\wedge dt.
\label{2form}
\end{equation}

Let us first
make an integration of the above equation for one period ($T$)
and then
average over the variable $\bar\phi_0$
.
Keeping in mind that the area meaning of the differential 2-form,
One  will find
immediately that the term  bracketed in  the expression of the geometric
phase Eq.(\ref{Fgphase}) corresponds to the left  hand side of the 
above equation, whereas the right hand side  will equal to $n\hbar$ times
the classical Hanny's angle (see Eq. (\ref{Hannay})).
The $\frac 1 2$ relation between the last term in Eq.(\ref{Fgphase}) and the 
classical angle is given by Ge and Child\cite{GC97} and verified by our 
explicit perturbative
results in follows. Then, we  can  reach  a simple relation between 
the geometric phase and non-adiabatic Hannay angle, 

\begin{equation}
\lambda_G^R=-(n+\frac 1 2)\Theta_H .
\label{Gphagle}
\end{equation}

Now  we  take a specific  
choice of 
the periodic parameters as an example to demonstrate the above approach
 and verify our findings.
Set that $a(t)=1+\epsilon \cos(\omega t),
b(t)=1-\epsilon \cos(\omega t) , c(t)=\epsilon\sin(\omega t)$.        
Our discussions are restricted to the elliptic case,      
namely,  $a(t)b(t) > c^2(t)$, i.e.
$\epsilon < 1$. 
The perturbation method will be employed in the following discussions.
Our solutions of power series are accurate to second order.

Now,  we rewrite the classical Hamiltonian in terms
 of the action-angle variables, i.e.
 $q=\sqrt{2I}\sin\phi,\quad p=\sqrt{2I}\cos\phi$,

\begin{equation}
 H_{cl} = H_0 (I) + \epsilon H_1(I,\phi),
\label{Hcl}
\end{equation}
where $H_0 = I, H_1 = -I\cos(\omega t + 2\phi)$.
It is convenient to employ the Lie transformation\cite{LL83} method
to make a canonical transformation, so that the new Hamiltonian
$\bar H(\bar I)$ contains the action variable only,

\begin{equation}
\bar H(\bar I) = \bar I -\frac{\bar I}{\omega+2} \epsilon^2.
\label{HI}
\end{equation}

The generating functions are 
$w_1=I\sin (\omega t + 2 \phi)/(\omega +2)$ and
$w_2 = 0$, respectively. 
The relation between the old variables and the new variables is
given by
$(\phi, I)= {\cal T}^{-1} (\bar \phi, \bar I)$,
where the transformation operator ${\cal T}^{-1} = 1 + \epsilon {\cal L}_1 +
\epsilon^2 ({\cal L}_2/2+{\cal L}_1^2/2).
{\cal L}_n$ is Lie operator defined by ${\cal L}_n = [w_n,~].
[~,~]$ represents a Poission Bracket. 
With the help of Eqs. (\ref{Hannay}) and (\ref{Hcl},\ref{HI}) 
we  arrive at the expression of the classical angle analytically,
\begin{equation}
\Theta_H = \frac{2\pi\epsilon^2}{(\omega +2)^2}.
\label{Hannay2}
\end{equation}

Obviously,  this classical non-adiabatic Hannay's angle
is independent of the action.
T-periodic  solution $(G_p(t),\Pi_p(t))$ of the Hamiltonian $H_{fl}$,
can be  derived  by using the power-series expansion, 
\begin{equation}
 G_p(t)=\frac {1} {2} -\frac{\cos(\omega t)}{\omega+2}\epsilon,\qquad
 \Pi_p(t)=-\frac{\sin(\omega t)}{\omega+2}\epsilon.
\label{GPperiod}
 \end{equation}
Notice the fact that an  arbitrary 
$\omega$ can be approached by a series of rational number
like $q/p$, we can repeat the above process by constructing a state
as in Eq.(\ref{stater}), where the $r=q$\cite{LHL98}. Finally, 
 we  obtain the analytic expression of the geometric phase,
\begin{equation}
 \lambda_G^R=-(\frac{\bar I_0}{\hbar}+\frac 1 2)
\frac{2\pi\epsilon^2}{(\omega+2)^2}.
\end{equation}

Considering the quantization rule (${\bar I_0}=n\hbar$) and
the explicit expression of Hannay's angle (\ref{Hannay2}), the above equation
coincise with the relation (\ref{Gphagle}).

An interesting example is given by the case $n=0$, i.e. the ground state
of the Floquet states. It corresponds to a 
cyclic squeezed state with period $T$, whose expectation values keep fixed
at the zero point, while its fluctuations change periodically (see 
Eq.(\ref{GPperiod})). The geometric phase of this cyclic
state resulting only from  the periodic evolution of the  fluctuations' part, 
is equal to one half of the classical Hannay angle. This is just what 
obtained by Ge and Child\cite{GC97}.

In summary, the squeezed state approach is used to study the nonadiabatic
geometric phase relating to the cyclic evolutions of a generalized
harmonic oscillator.  The quantum phases are obtained explicitly and found
to be $-(n+1/2)$ times the Hannay angle. The quantum phase can be
interpreted as a sum of the area difference on the expectation value plane
through the cannonical transformation and the area on the quantum
fluctuation plane swept out by a periodic orbit. The explanation given
here provides a unified picture of the geometric meaning of the quantal
phase for the adiabatic case as well as the nonadiabatic case.  In the
adiabatic limit, our $n+1/2$ relation is identical to the elegant formula
of Berry\cite{HB85}. However, the semiclassical approximation has not been
envoked. 
\vspace{.3cm}\\ 
We would like to thank Profs. Shi-Gang Chen, Lei-Han Tang, and 
Wei-Mou Zheng for helpful discussions and comments.
This work was supported
in part  by  the grants from the Hong Kong Research Grants Council (RGC)  
and the Hong Kong Baptist University Faculty Research Grants (FRG).

\end{multicols}
\end{document}